\documentclass[conference]{IEEEtran}
\IEEEoverridecommandlockouts
\usepackage[numbers]{natbib}
\setcitestyle{open={[},close={]}}

\usepackage{amsmath,amssymb,amsfonts}

\usepackage{graphicx}
\usepackage{textcomp}
\usepackage{xcolor}
\usepackage{float}
\usepackage{array}
\usepackage[font=footnotesize,labelfont=bf]{caption}
\usepackage{optidef}
\usepackage{subcaption}
\usepackage[utf8]{inputenc}
\def\BibTeX{{\rm B\kern-.05em{\sc i\kern-.025em b}\kern-.08em
    T\kern-.1667em\lower.7ex\hbox{E}\kern-.125emX}}

\usepackage{algorithm}
\usepackage{algpseudocode}
\usepackage{color, xcolor, colortbl}

\begin{document}

\title{Towards a Secure and Reliable Federated Learning using Blockchain\\}
\author{\IEEEauthorblockN{
Hajar Moudoud \IEEEauthorrefmark{1}\IEEEauthorrefmark{3},
Soumaya Cherkaoui  \IEEEauthorrefmark{1},
Lyes Khoukhi \IEEEauthorrefmark{2}
}

\IEEEauthorblockA{
\IEEEauthorrefmark{1} Department of Electrical and Computer Engineering, Université de Sherbrooke, Canada\\ 
\IEEEauthorrefmark{2} GREYC CNRS, ENSICAEN, Normandie University, France\\
\IEEEauthorrefmark{3} University of Technology of Troyes, France \\ 
\{hajar.moudoud, soumaya.cherkaoui\}@usherbrooke.ca, lyes.khoukhi@ensicaen.fr}
}

\maketitle
\begin{abstract}
Federated learning (FL) is a distributed machine learning (ML) technique that enables collaborative training in which devices perform learning using a local dataset while preserving their privacy. This technique ensures privacy, communication efficiency, and resource conservation. Despite these advantages, FL still suffers from several challenges related to reliability ($i.e. ,$ unreliable participating devices in training), tractability ($i.e. ,$ a large number of trained models), and anonymity. To address these issues, we propose a secure and trustworthy blockchain framework (SRB-FL) tailored to FL, which uses blockchain features to enable collaborative model training in a fully distributed and trustworthy manner. In particular, we design a secure FL based on the blockchain sharding that ensures data reliability, scalability, and trustworthiness. In addition, we introduce an incentive mechanism to improve the reliability of FL devices using subjective multi-weight logic. The results show that our proposed SRB-FL framework is efficient and scalable, making it a promising and suitable solution for federated learning.

\end{abstract}

\IEEEpeerreviewmaketitle

\section{Introduction}
\label{sec: Introduction}
The adoption of machine learning (ML) is expected to enable the flourishing development of many applications, such as autonomous driving, intelligent applications, or load balancing. Although machine learning has significantly improved the performance of many applications, most machine learning models must be built in central servers, which may violate privacy, especially for users with sensitive data \cite{6r,globecom1, globecom6}. This is because to participate in learning a model, users/devices trade their privacy by sending their sensitive data to the central server \cite{globecom13}. To overcome this problem, Google proposed Federated Learning (FL) to locally train models on edge devices while keeping their data private \cite{r14, globecom7}. Federated learning works as a collaborative learning process in which edge devices run training models locally and then send model updates, $i.e.$, the weight and gradient parameters of the training model, to the central server. 

Although FL has improved privacy protection, it still suffers from several problems. First, all devices involved in the FL process are expected to contribute their resources unconditionally, which is not accepted by all workers \cite{globecom8, globecom9}. Without reward, only a small fraction of the devices are willing to participate in the training process. On the other hand, the devices involved in the training are unreliable and may act maliciously, intentionally, or unintentionally, which may affect the overall model and lead to erroneous model updates \cite{globecom14}. For intentional behaviors, the device may send an incorrect update that influences the overall model and leads to failure. In addition, communication is a critical bottleneck in FL networks, which can lead to some unintended device behaviors. Due to the high communication turns between devices and the central server, devices may unintentionally update some poor parameters that negatively influence federated learning tasks. Therefore, it is important to develop a new FL architecture to alleviate transmission loads and to improve the reliability of model updates.

To overcome the aforementioned challenges, several researchers proposed combining FL and blockchain technology. Blockchain is a distributed solution that ensures trust between unreliable entities without depending on a third party \cite{3r, globecom15}. With its features of anonymity, reliability, security, and no single point of failure, blockchain has the potential to improve FL. The authors of the paper \cite{r1} proposed a FL and blockchain architecture, in which devices involved in learning send their model updates to the blockchain for validation. The work in \cite{r2} proposed using the blockchain with FL to establish a secure reputation for devices participating in learning. While this work presented the potential of using blockchain with FL to improve device reputation and trustworthiness, it introduced third-party devices, namely blockchain miners, to mine blocks in a distributed manner. This will lead to model leakage, as the model parameters will be shared with these miners. In addition, blockchain still suffers from several shortcomings, including the scalability problem, which can affect its integration with FL. For example, the scalability problem of the blockchain may further increase the communication delay for the FL. 

To address these challenges, we present in this paper the SRB-FL framework. SRB-FL provides secure, reliable, trustworthy, and transparent framework suitable for FL. This framework implements: (1) blockchain sharding to enable parallel model training to improve the scalability of blockchain FL; and (2) an incentive mechanism to improve the reliability of FL devices. We consider synchronous FL case where devices participating in the learning belong to the same blockchain shard, that is to enable all devices to follow the same time line.

The main contributions of our paper can be summarized as follows:
\begin{itemize}
    \item We design a blockchain framework, called SRB-FL, which is a secure and reliable framework that is suitable for FL. SRB-FL uses an incentive mechanism to improve the reliability of FL devices and leverages blockchain sharding to ensure a scalable training.
    \item We propose a reputation mechanism that uses subjective logic (SL) to incite FL devices to provide reliable model updates.
    \item We evaluate the performance of the SRB-FL framework in terms of efficiency, latency, reputation, and scalability. The experimental results show that SRB-FL can effectively guarantee reliability, security, and scalability, making it a promising solution for securing FL.
\end{itemize}
The rest of this paper is organized as follows. Section II presents a review of some related work. Section III describes the SRB-FL framework for FL and explains its main components. Section IV defines the proposed incentive mechanism. Section V presents the performance evaluation of SRB-FL. Finally, Section VI concludes the paper.

\section{Related work}
\label{sec: Related works}
Federated learning was introduced by Google to distribute data learning, enabling FL devices to learn locally without sharing their sensitive data. Keeping data locally yields many benefits, namely immediate access to ephemeral data, privacy preservation and leveraging the computing resources of the sparse participants \cite{globecom2, globecom5, globecom10, globecom11}. However, FL still suffers from several shortcomings such as device’s reliability and security \cite{5r}. To overcome some of these problems, research has introduced blockchain-based FL \cite{r4,r5, globecom12}. In \cite{r6}, the authors proposed DeepChain, a secure and reliable FL framework that improves the reliability of FL devices using blockchain smart contracts. This framework implemented an incentive mechanism to incite FL devices to behave honestly. However, a comprehensive study with real-time data is needed to scale up the DeepChain approach. In \cite{r7}, the authors proposed a trustworthy blockchain and FL solution to ensure reliable shard training on the fog. Although useful, their solution has high a throughput.

In \cite{r11}, the authors proposed a blockchain-based on-device FL architecture, in which local and global updates of FL devices are stored on the blockchain. They also studied the scalability, robustness, and latency of integrating blockchain with FL. In \cite{r12}, the authors presented a FL and blockchain model that improves the security. In this model, the FL devices upload only the final updates on the chain, whereas the local updates are sent to a distributed hash table. Although this model reduces block generation latency, it is subject to security and integrity issues. Indeed, the authors proposed to reduce the consensus difficulty of the blockchain in order to reduce the latency, which can consequently lead to false data injection attacks or 51\% majority attacks. 

In \cite{r8}, the authors proposed a reliable FL device selection, where only reliable devices participate in the model training. This solution used a consortium blockchain approach to improve the device reputation. In \cite{r9}, the authors proposed a hybrid blockchain architecture for a secure and reliable federated learning. This architecture proposed integrating learned models into the blockchain to improve the reliability of data. Although the combination of Blockchain with FL improved to be very useful, however, to the best our knowledge, none of the existing works have taken into considered the inherent problems of blockchain, such as the scalability problem. Indeed, blockchain suffers from issues related to the block generation which impact the scalability of blockchain.

To address some of the shortcomings of these existing solutions, we propose the SRB-FL framework. SRB-FL provides a secure and reliable blockchain framework that allows FL devices to train and send model updates in a secure, reliable, and trustworthy manner. The SRB-FL framework proposes the use of multiple shard chains, each of which is responsible for training a FL model, and the main chain is responsible for tracing all final model updates. In addition, we propose to use an incentive mechanism based on subjective logic to improve the reliability of FL devices.

\section{System Model}
\label{sec: system}
The SRB-FL is a secure and reliable blockchain FL framework that enables FL devices to train and send model updates securely. In this section, we first describe SRB-FL’s framework. Then, we explain how the blockchain sharding and incentive mechanism are carried out.
\subsection{SRB-FL Framework}
We consider the synchronous case, where FL devices can train and send updates at the same time. For this purpose, we propose to use blockchain sharding to allow several models to use the blockchain synchronously. Each model is assigned to a group of devices belonging to the same shard (or cluster) to train and send local updates. Once these devices have completed their training using their local data, they send their final model updates to the main chain. Consequently, several models can train and use the blockchain efficiently.

Fig. 1 shows a blockchain and FL framework, including both a model layer and a sharding layer. For the model layer, we consider a network consisting of edge devices that are connected by a wireless network ($i.e.,$ LTE, 4G, 5G, etc.). These devices have sufficient computational power to train the models and communicate their updates. In the context of FL, these devices have private data with which they will train the model. Therefore, these devices can be considered as FL workers/devices. Each FL device can train a shared model using its local data and generate model updates. Then, all FL devices send their model updates to their allocated shard. This process is repeated until the desired model accuracy is achieved.
\begin{figure*}[t]
	\centering
	\includegraphics[scale=0.4]{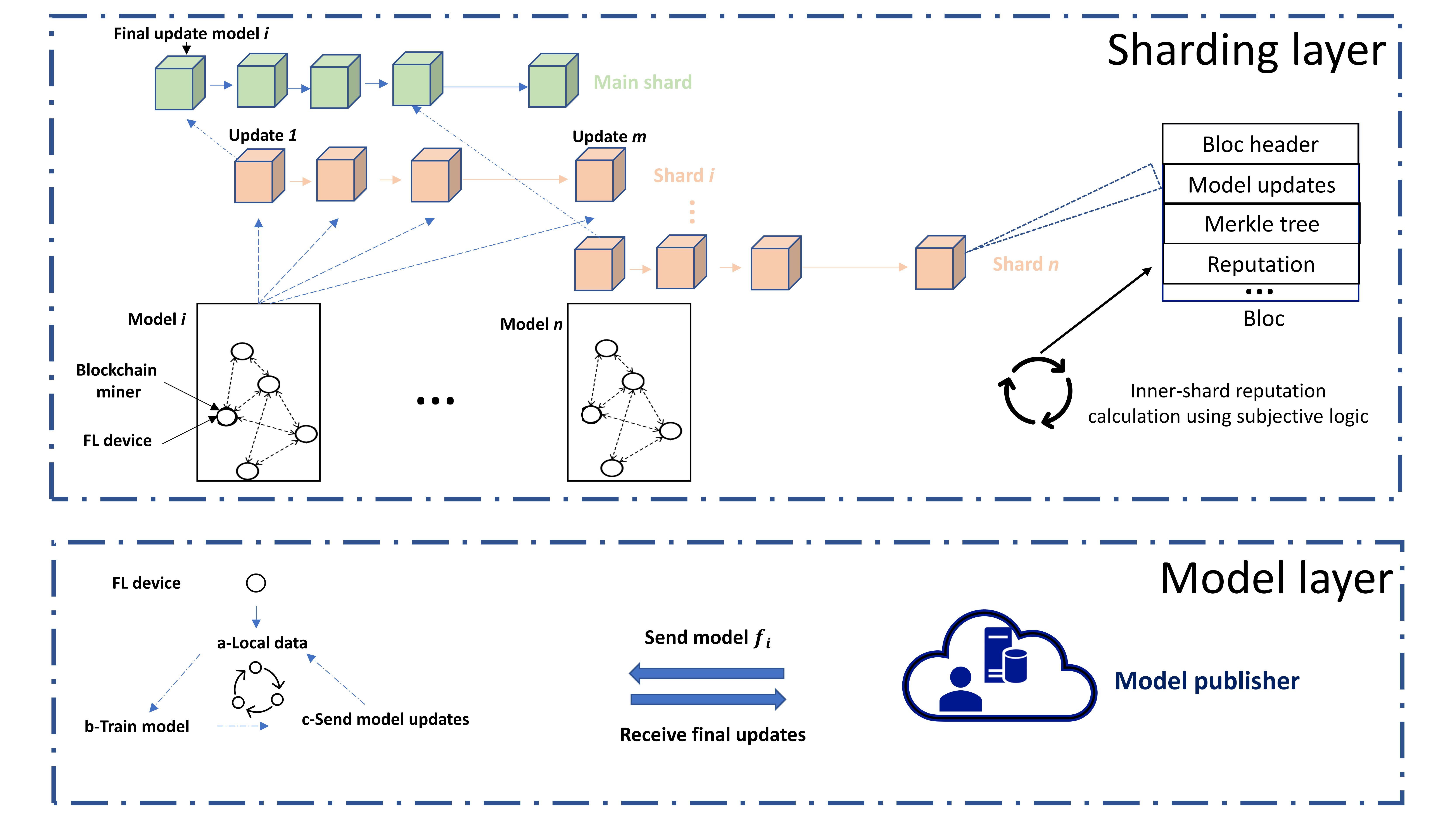}
	\caption{System model of reliable blockchain and FL using sharding.}
	\label{fig:archi}
\end{figure*}

In the sharding layer, FL devices act as blockchain miners because they have sufficient computing power. The sharding layer consists of the main chain and the sub-chains. The main chain stores final model updates, while the sub-chains store local model updates.  The sub-chains act as a group of clusters, each sub-chain contains a group of FL devices. The details of device selection for each sub-chain are not the focus of this paper. However, we can adopt some of the well-known methods, such as: feature selection, random selection or reputation-based selection.
\subsection{Framework Components}
We consider that our blockchain sharding layer can support multiple learning models at the same time in a fully distributed manner. In this case, a blockchain transaction $T_{j,i} (t)$ can be viewed as a model update at time $t$ sent by a FL device. 

We design a reliable and scalable blockchain suitable with FL, we use blockchain smart contracts to implement an incentive mechanism that improves the reliability of FL. The incentive mechanism includes five steps as follows:
\begin{itemize}
    \item Step 1 (\textit{ Initialization}): Model publisher creates a smart contract with the FL model and specific requirements ($e.g.,$ data type, accuracy, training time, etc.). The model is then affected to the shard with devices that satisfies some of these requirements.
    \item Step 2 (\text {Training}): Beginning of local training, where each FL devices $\it{j,k}$ belonging to a shard $j$ has a local dataset $D_k$ with a data size of $\left | D_{k} \right |$. The FL aims to find the optimal global model parameters $\boldsymbol{w}\in \mathbb{R}^{l}$ that minimize the global loss function $f(\boldsymbol{w})$:
\begin{equation}
    \min_{\boldsymbol{w}\in \mathbb{R}^{l}} f(\boldsymbol{w}) = \frac{1}{D} \sum_{k=1}^{N} f_k(\boldsymbol{w}), 
\end{equation}
where $D$ is the total data sample from $N$ FL device ( $D = \sum_{k=1}^{N}\left | D_{k} \right |$) and $f_k()$ is the local loss function computed by a model device based on its local training data.

\item Step 3 (\textit{Updates}): After the training, the FL devices send updates to the corresponding sub-chain. It is worth mentioning that only the final updates that correspond to the highest accuracy are added into the main chain. To add a block into the chain, a FL device acts as miners where they need to reach a consensus. Once the consensus is reached among the shard, the final model updates are then stored permanently into the chain. In our framework, we propose using on-chain and off-chain storage, that is, metadata corresponding to the updates are stored on the chain, and the raw data are stored off the chain. This is because the blockchain has limited bloc size. Finally, the updates are sent to the corresponding model publisher.

\item Step 4 (\textit{Reputation}): To improve the reliability of devices, we use an inter-shard reputation process based on subjective logic, where devices that performed honestly are rewarded and devices performing malicious are penalized, respectively. 

\end{itemize}

\section{Secure and reliable blockchain-FL using Multiweight Subjective Logic}
\label{sec:proposed}
In order to improve the security and reliability of blockchain-based FL, two mechanisms are proposed: (1) a secure sharding process using subjective logic (SSSL) to improve the reliability of FL devices, and (2) a lightweight sharding consensus (LWSC) to improve the security of FL.
\subsection{Key Notations and Assumptions}
Let $O_i^j$ denotes the subjective opinion or trust of a device $i$ and $j$ is the target to which the opinion applies ($i.e.,$ model updates), where $O= \{b,d,u,a\}$ and $b$, $d$, and $u$ respectively represent the belief, disbelief, and uncertainty of $i$, $b+d+u=1$ and $b$, $d$, and $u \in [0,1]$. The parameter $a$ refers to the base rate, it is used to compute the opinion probability expectation value $E(O)=b+au$. Each shard $S$ trains a FL model, where it holds the vector $O_{i,t}^j$ ($i.e.,$ trust in a FL device) during an epoch $t$ with $a$ uncertainty. In the subjective logic, there can be different opinions ($e.g.,$ $ hyper$ $opinions$, $multinomial$ $opinions$, and $binomial$ $opinions$). In this article, we are interested in uncertain binomial opinion (UBO) over binary events ($i.e.,$ valid or invalid update). This is because, in the FL context, the uncertainty about an update is $u \geq 0$. Let $Beta (X, \alpha,\beta)$ corresponds to a beta probability density function \cite{r13}, where
$
\alpha= r+2a$ and
$\beta=  s+2(1-a) 
$. Let $r$ denotes the number of observations of $X$, and $s$ denotes the number of observations of $\bar{X}$.

We assume that each shard can validate one or several updates. Each shard is composed of multiple FL devices that could be either dishonest or honest. These devices are responsible for validating the updates by reaching a consensus. A blockchain consensus is a fault-tolerant mechanism used to reach an agreement between all the devices. Furthermore, we assume that at the beginning, all the shards have equal rate probability $a$ and opinion $O_{i,t}^j$, respectively. In this article, to obtain accurate reputation values of devices, every shard combines its intra-trust opinion with inter-trust option to generate the composite trust value for devices.   
\subsection{Secure Sharding Using Subjective Logic}
Subjective Logic (SL) is one of the prominent probabilistic logic that evaluates trustworthiness of different entities ($i.e.,$ miners/ Fl devices). The SL uses opinions about probability values to determine a subjective belief through belief, disbelief, and uncertainty statements. 
During a time epoch $te$ different from $t$ ($t$ is a fixed value by the model publisher whereas $te$ is a fixed value by the shard devices), the reputation of a FL devices $i$ regarding an update $t$ in a time slot can be expressed as follows: 
\begin {equation}
O_{i \to t}^{te}= \{b_{i \to t}^{te},d_{i \to t}^{te}, u_{i \to t}^{te}\}
\end{equation} 
Based on SL and UBO, we can obtain:

\begin{equation}
\begin{cases}
b_{i \to t}^{te}= \frac{\alpha_i^{te}}{\alpha_i^{te}+\beta_i^{te}} \\
d_{i \to t}^{te}= \frac{\beta_i^{te}+2a-2}{\alpha_i^{te}+\beta_i^{te}} \\ 
u_{i \to t}^{te}= \frac{2}{\alpha_i^{te}+\beta_i^{te}} \\
a= \text{base rate of X}
\end{cases}
\end{equation} 
Note that a dogmatic opinion of the reliability of a device with $u=0$ is equivalent to a defined probability density function ($i.e.,$ trustworthy or untrustworthy). The shard treats a FL device as reliable after $r$ observations of $X$ and $s$ observations of $\bar{X}$ with a base rate $a$. According to the Beta probability density function, the value of a FL device's opinion (trust) $i$ regarding the update $X$ during an epoch $te$ is defined as follows:  
\begin {equation}
E(O_{i \to X}^{te})=b_{i \to X}^{te}+au_{i \to X}^{te}
\end{equation}
Shard verification process can be expressed as follows: 
\begin{equation}
\
V_j(t)_=\sum_{i=1}^{n} V(t_i)E(O_{i \to t_i}^{te})
\end{equation}
where, $n$ is the number of active FL devices in a shard $j$. Every verification is weighted after being collected from a shard to the main chain for accurate reputation update. The weight of a verification is set based on multiple factors. Consequently, the traditional SL is evolved toward multi-weighted SL to consider different factors. 

To ensure a secure sharding, we expect to focus on devices with higher reputation values to train a FL model while avoiding being misled by malicious devices.
Note that the reputation of a FL device is affected by many factors among them we find interactivity and novelty. Here, we consider the following factors to calculate the trustworthiness:
\begin{itemize}
    \item Interactivity: Inside a shard, devices could either interact in a honest or dishonest manner. The honest interactions increase the trust of the device, which leads to increase the reward, and vice versa. To encourage honest behavior of FL devices, honest devices have a higher weight in the reputation process. Similarly to work \cite{r2}, the interactivity value of a device $i$ in a shard $j$ during a time epoch $te $, $S_{i}^{te}$, can be written as:
\begin{equation}
S_i^{te}=b_i^{te}+ d_i^{te}=1-u_i^{te}
\end{equation}
     \item Novelty: The trustworthiness of a FL device changes with time, the novelty is defined to measure the freshness of devices interaction inside a shard. Recent device interactions with more novelty have a higher reward than past interactions. To reflect the effect of time on reputation opinion, we define the novelty function as follows:
\begin{equation}
n_i^{te}=\nu(tn_i-te)^{-\mu}
\end{equation}
where $\nu$ and $\mu$ are two pre-defined parameters to adapt the effect of novelty, chosen similar to work in \cite{r2}. 
\end{itemize}
To sum up, by considering the interactivity and novelty, the overall weight of the reputation opinion $W_i^{te}$ of a FL device $i$ is calculated by:
\begin{equation}
f_i^{te}=w_1 n_i^{te} + w_2 S_i^{te}
\end{equation}
where $w_1$ and $w_2$ are pre-defined weight factors that satisfy $w_1 + w_2 =1$. Consequently, the reputation opinion of a FL device $i$ is aggregated with the weights factors to acquire an intra-reputation opinion denoted $w_{i,j}^{intra} = (b_{i,j}^{intra}, d_{i,j}^{intra}, u_{i,j}^{intra}, a_{i,j}^{intra})$, for a device $i$ inside shard $j$ during a time window $te\leq n$. The inter-reputation opinion is equal to the weighted reputation average of all the collected reputation opinions. The inter-reputation opinions refer to opinions that are formed after observing a targeted miner behavior. Therefore, the intra-reputation opinions of a device $i$ in a time epoch $te$ are denoted as:

\begin{equation}
\begin{cases}
b_{i,j}^{intra}= \frac{\sum_{i=1}^n f_i^{te} b_i^{te}}{\sum_{i=1}^n f_i^{te}}\\
d_{i,j}^{intra}= \frac{\sum_{i=1}^n f_i^{te} d_i^{te}}{\sum_{i=1}^n f_i^{te}}\\
u_{i,j}^{intra}= \frac{\sum_{i=1}^n f_i^{te} u_i^{te}}{\sum_{i=1}^n f_i^{te}}\\
\end{cases}
\end{equation}
and 
$E(O_{i}^{te})^{intra}=  \frac{\sum_{i=1}^n f_i^{te} E(O_i^{te}) }{\sum_{i=1}^n f_i^{te}}$ 

\subsection{Inter and Intra Reputation Process}
\begin{figure}[t]
	\centering
	\includegraphics[width=9cm, height=5cm]{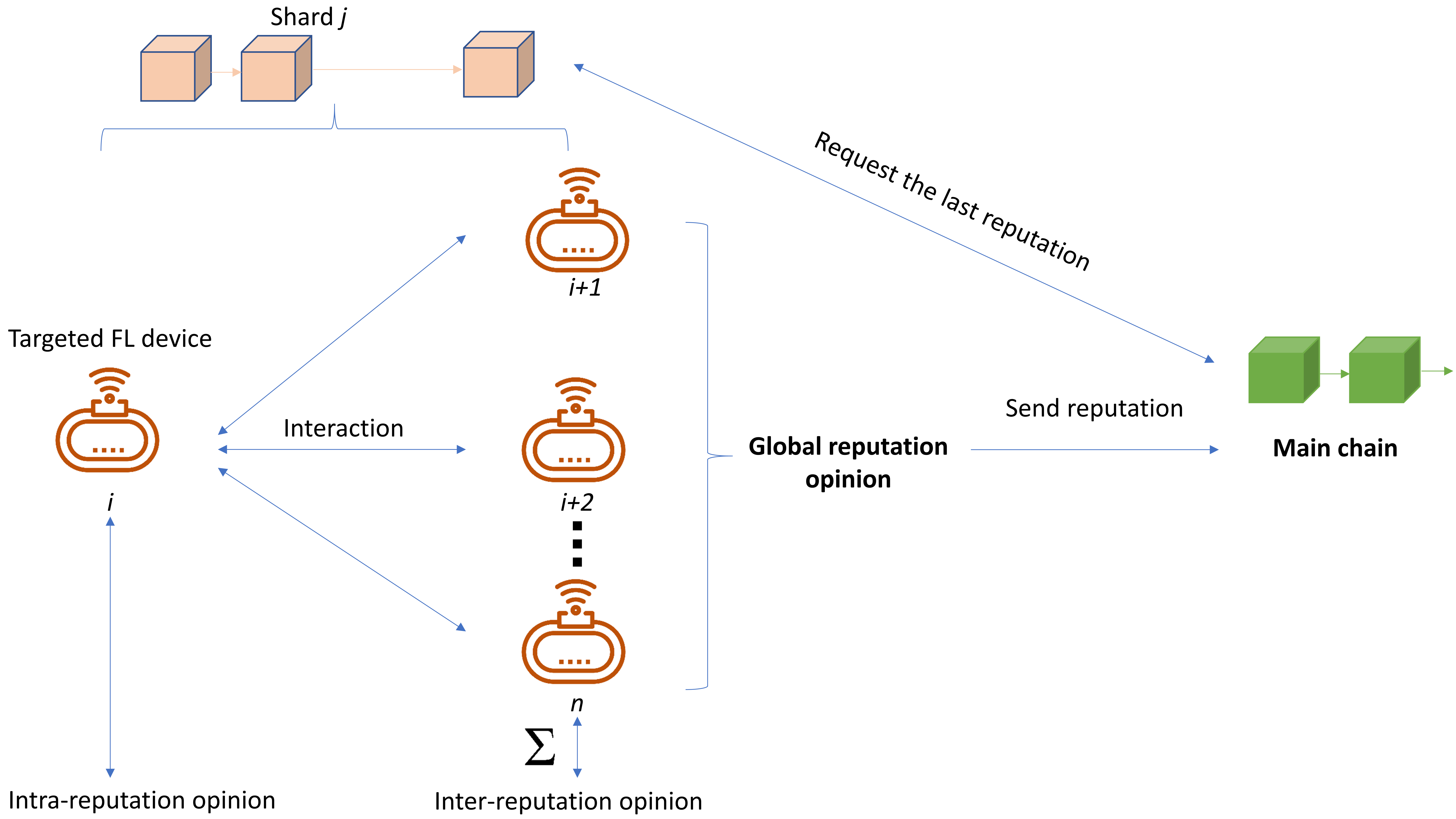}
	\caption{The operation of inter and intra reputation process.}
	\label{fig:archi}
\end{figure}
\begin{table}[t]
	
	\begin{center}
		
		\caption{SIMULATION SETUP PARAMETER}\label{tab:2}
		
		\begin{tabular}{c | c}
			\hline
			\hline
			\textbf{Parameters} & \textbf{Value}\\
			\hline
			Classifier 
			Training examples & 1000\\
			Weight parameters  &  $\alpha$=0.4;$\beta$=0.6\\
			Epoch  & 10\\
			IP & localhost\\
			Shards &  8\\
			FL devices & 10 \\        
			PoW difficulty & 4\\
			Predefined parameters & $w_1$=$w_2$=$a$=0.5; te=60min; u=1\\
			\hline
			\hline
		\end{tabular}
		
	\end{center}
\end{table} 
The operation of the inner shard reputation aims to utilize local knowledge on the behavior of a Fl device with the other devices inside the shard. Generally, inner reputation is based on collecting reputation for centralized analysis. Consequently, we define intra-reputation and inter-reputation opinions. The intra-reputation opinion of a device $i$ is aggregated with the weight factors. As for the inter-reputation opinion is formed by other devices after observing the targeted device. To explain the basic concepts, we present a toy example of a reliable blockchain sharding for IoT systems in Fig. 2. In this example, a device $i$ has an intra-reputation opinion about an update $u$. First, the neighbor devices interact with the targeted device. Then, they form an individual opinion about it. Subsequently, they forward these opinions to the shard chain. According to the interactivity and novelty, the inter-reputation opinions are weighted. Finally, the global reputation which is equal to the joint combination of the inter-reputation and intra-reputation is updated to the main chain. If the targeted device is a new FL device inside the shard, the FL device sends a private key to the shard to be authenticated. The inter-reputation opinion $E(O_{i}^{te})^{inter}$ is calculated using other devices opinions regarding the trustworthiness of miner $i$.

\section{Evaluation}
\label{sec: evaluation}
This section presents the evaluation of SRB-FL framework. We use MNIST data and a widely used software environment PyTorch to perform a digit classification task to evaluate the proposed framework. The simulations were conducted on a laptop with 2.2 GHz Intel i7 processor and 16 GB of memory \cite{globecom4, globecom3}. Simulation parameters are illustrated in Table 1.

\begin{figure*}[t]
	\centering
	\includegraphics[scale=0.4]{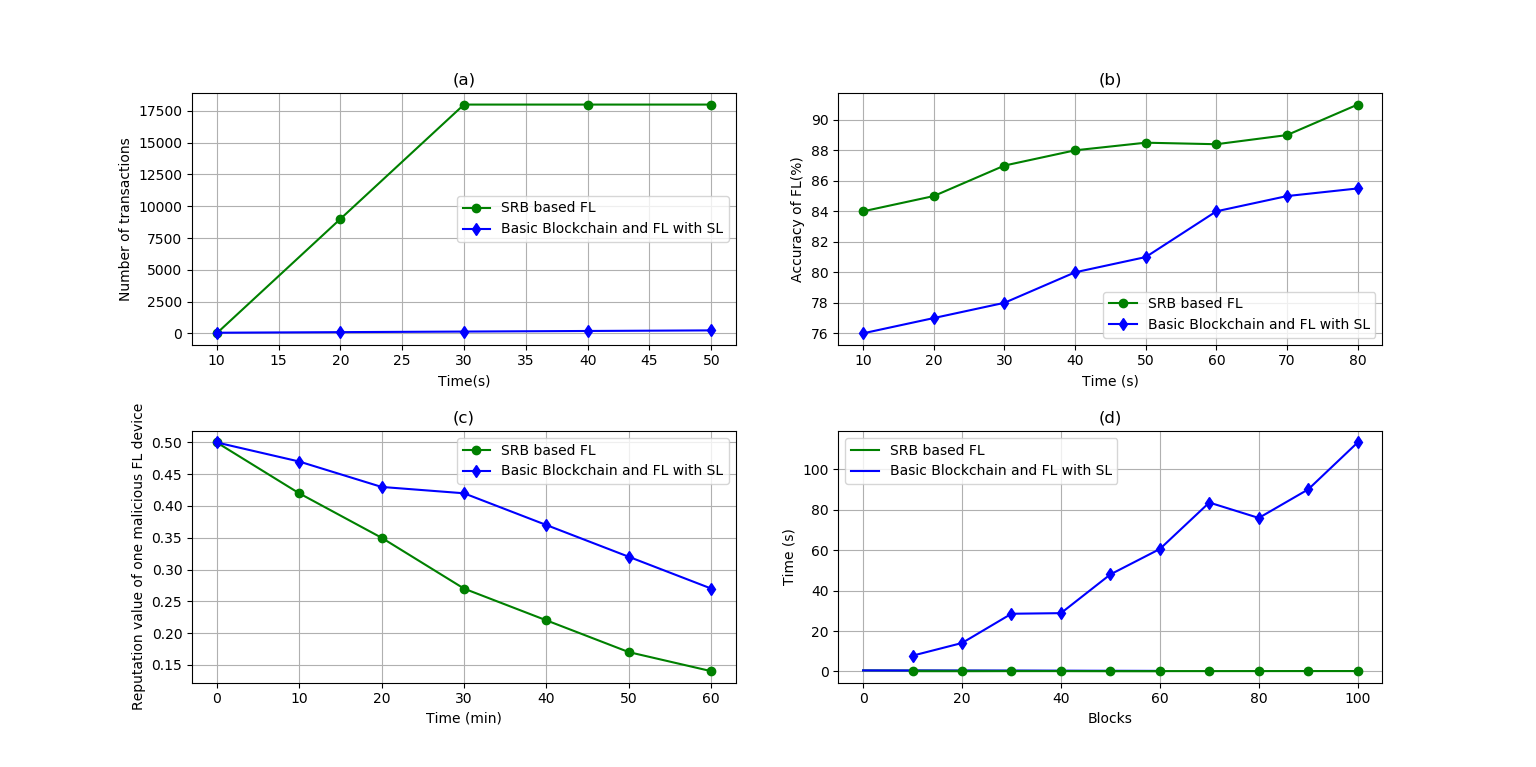}
	\caption{Performance evaluation of SRB based FL in comparison with a basic blockchain framework in terms of: a) Scalability; b) FL Accuracy; c) Reputation; and d) Latency.}
	\label{fig:archi}
\end{figure*}

We compare the performances of our SRB-FL and a basic blockchain-FL framework similar to the work \cite{r2}. This framework uses blockchain technology, specifically Proof-of-Work (PoW) consensus and subjective logic to ensure a reliability of FL devices. Fig. 3 shows the performance evaluation of SRB-FL in terms of: a) Scalability; b) FL Accuracy; c) Reputation; and d) Latency.

Fig. 3(a) examines the scalability of the two frameworks, we evaluate the number of transactions ($i.e.,$ number of model updates) in a function of time. Because SRB based FL uses the sharding concept, we can observe that the number of transactions in a function time does not increase drastically, in the opposite of a basic blockchain-FL framework. This because the basic framework uses Proof-of-Work (PoW) consensus, which a high-computing consensus algorithm that limits the number of transactions/time. We clearly observe that the proposed SRB-FL framework is more scalable. 

Fig. 3(b) evaluates the effects of the unreliable FL devices on accuracy of federated learning. In this evaluation, we increase the number of unreliable workers in a function of time and observe their effect on the FL accuracy. We can observe that the increasing number of unreliable FL devices can bring decreasing federated learning accuracy. For example, a blockchain based FL framework with only reliable workers achieves an accuracy of 96\%, while the accuracy of the SRB-FL framework achieves 94\% and a basic blockchain based FL framework achieves 84.6 \% accuracy, which is 1.4\% lower. 

Fig. 3(c) evaluates the reputation value of malicious FL devices that send erroneous updates. As illustrated, the use of inter and inter reputation based on multiweight subjective logic improves the reputation in comparison with only a reputation mechanism that uses subjective logic. Indeed, the reputation of SRB-FL is more accurate than the reputation used in a basic blockchain-FL framework. The reputation value of a malicious FL devices are calculated during a time frame of 60 $min$ and lowers to 0.14, which is lower than the reputation calculated using subjective logic (0.26). 

Fig 3(d) compares the latency of variant blockchains-FL. As shown by the figure, the basic blockchain-FL uses PoW consensus, which increased latency when increasing the number of blocks. Meanwhile, the time needed to add a block to the blockchain is a fixed value 0.1s for the SRB-FL. The results demonstrate the efficiency of SRB-FL, in terms of scalability, accuracy, reputation, and latency in comparison with a basic blockchain-FL framework.

\section{Conclusion}
\label{sec: Conclusion}
This paper proposed SRB-FL, a secure and reliable blockchain framework suitable for federated learning. To overcome the issues of unreliable and malicious FL devices involved in the training of FL that may affect the overall model accuracy and lead to falsified model updates, we proposed a secure and reliable blockchain framework. We employed blockchain sharding to improve blockchain efficiency and enable parallel model training. We then designed a reputation mechanism that uses a multi-weight subjective logic to improve the reliability FL devices. Furthermore, we designed an intra-reputation and inter-reputation process to simulate the reputation of FL devices that have trustworthy and reliable updates to participate on the learning process. The simulation results showed that SRB-FL framework can greatly improve the reliability of FL devices while ensure reliable federated learning making it a promising solution to improve the reliability and security of federated learning.

\section*{acknowledgement}
\label{sec:acknowledgement}
The authors would like to thank the Natural Sciences and Engineering Research Council of Canada, as well as FEDER and GrandEst Region in France, for the financial support of this research.

\label{Related work}
\bibliographystyle{IEEEtran}
\bibliography{./references}

\begin{thebibliography}{10}
\providecommand{\url}[1]{#1}
\csname url@samestyle\endcsname
\providecommand{\newblock}{\relax}
\providecommand{\bibinfo}[2]{#2}
\providecommand{\BIBentrySTDinterwordspacing}{\spaceskip=0pt\relax}
\providecommand{\BIBentryALTinterwordstretchfactor}{4}
\providecommand{\BIBentryALTinterwordspacing}{\spaceskip=\fontdimen2\font plus
\BIBentryALTinterwordstretchfactor\fontdimen3\font minus
  \fontdimen4\font\relax}
\providecommand{\BIBforeignlanguage}[2]{{%
\expandafter\ifx\csname l@#1\endcsname\relax
\typeout{** WARNING: IEEEtran.bst: No hyphenation pattern has been}%
\typeout{** loaded for the language `#1'. Using the pattern for}%
\typeout{** the default language instead.}%
\else
\language=\csname l@#1\endcsname
\fi
#2}}
\providecommand{\BIBdecl}{\relax}
\BIBdecl

\bibitem{6r}
Z.~Abou El~Houda, A.~S. Hafid, and L.~Khoukhi, ``Cochain-{SC}: {An} {Intra}-
  and {Inter}-{Domain} {Ddos} {Mitigation} {Scheme} {Based} on {Blockchain}
  {Using} {SDN} and {Smart} {Contract},'' \emph{IEEE Access}, vol.~7, pp.
  98\,893--98\,907, 2019.

\bibitem{globecom1}
E.~D. Ngangue~Ndih and S.~Cherkaoui, ``On {Enhancing} {Technology}
  {Coexistence} in the {IoT} {Era}: {ZigBee} and 802.11 {Case},'' \emph{IEEE
  Access}, vol.~4, pp. 1835--1844, 2016.

\bibitem{globecom6}
Z.~Mlika and S.~Cherkaoui, ``Competitive {Algorithms} and {Reinforcement}
  {Learning} for {NOMA} in {IoT} {Networks},'' in \emph{{ICC} 2021 - {IEEE}
  {International} {Conference} on {Communications}}, Jun. 2021, pp. 1--6.

\bibitem{globecom13}
H.~Moudoud, L.~Khoukhi, and S.~Cherkaoui, ``Prediction and {Detection} of
  {FDIA} and {DDoS} {Attacks} in {5G} {Enabled} {IoT},'' \emph{IEEE Network},
  vol.~35, no.~2, pp. 194--201, Mar. 2021.

\bibitem{r14}
J.~Konečný, H.~B. McMahan, D.~Ramage, and P.~Richtárik, ``Federated
  {Optimization}: {Distributed} {Machine} {Learning} for {On}-{Device}
  {Intelligence},'' \emph{arXiv:1610.02527 [cs]}, Oct. 2016.

\bibitem{globecom7}
A.~Taïk, H.~Moudoud, and S.~Cherkaoui, ``Data-{Quality} {Based} {Scheduling}
  for {Federated} {Edge} {Learning},'' in \emph{2021 {IEEE} 46th {Conference}
  on {Local} {Computer} {Networks} ({LCN})}, Oct. 2021, pp. 17--23.

\bibitem{globecom8}
A.~Taïk and S.~Cherkaoui, ``Electrical {Load} {Forecasting} {Using} {Edge}
  {Computing} and {Federated} {Learning},'' in \emph{{ICC} 2020 - 2020 {IEEE}
  {International} {Conference} on {Communications} ({ICC})}, Jun. 2020, pp.
  1--6.

\bibitem{globecom9}
A.~Tak and S.~Cherkaoui, ``Federated {Edge} {Learning}: {Design} {Issues} and
  {Challenges},'' \emph{IEEE Network}, vol.~35, no.~2, pp. 252--258, Mar. 2021.

\bibitem{globecom14}
H.~Moudoud, S.~Cherkaoui, and L.~Khoukhi, ``Towards a {Scalable} and
  {Trustworthy} {Blockchain}: {IoT} {Use} {Case},'' in \emph{{ICC} 2021 -
  {IEEE} {International} {Conference} on {Communications}}, Jun. 2021, pp.
  1--6.

\bibitem{3r}
Z.~A. El~Houda, A.~S. Hafid, and L.~Khoukhi, ``Blockchain-based {Reverse}
  {Auction} for {V2V} charging in smart grid environment,'' in \emph{{ICC} 2021
  - {IEEE} {International} {Conference} on {Communications}}, Jun. 2021.

\bibitem{globecom15}
H.~Moudoud, S.~Cherkaoui, and L.~Khoukhi, ``An {IoT} {Blockchain}
  {Architecture} {Using} {Oracles} and {Smart} {Contracts}: the {Use}-{Case} of
  a {Food} {Supply} {Chain},'' in \emph{2019 {IEEE} 30th {Annual}
  {International} {Symposium} on {Personal}, {Indoor} and {Mobile} {Radio}
  {Communications} ({PIMRC})}, Sep. 2019, pp. 1--6.

\bibitem{r1}
H.~Kim, J.~Park, M.~Bennis, and S.-L. Kim, \emph{On-{Device} {Federated}
  {Learning} via {Blockchain} and its {Latency} {Analysis}}, Aug. 2018.

\bibitem{r2}
J.~Kang, Z.~Xiong, D.~Niyato, S.~Xie, and J.~Zhang, ``Incentive {Mechanism} for
  {Reliable} {Federated} {Learning}: {A} {Joint} {Optimization} {Approach} to
  {Combining} {Reputation} and {Contract} {Theory},'' \emph{IEEE Internet of
  Things Journal}, vol.~6, no.~6, pp. 10\,700--10\,714, Dec. 2019.

\bibitem{globecom2}
A.~Rachedi, M.~H. Rehmani, S.~Cherkaoui, and J.~J. P.~C. Rodrigues, ``{IEEE}
  {Access} {Special} {Section} {Editorial}: {The} {Plethora} of {Research} in
  {Internet} of {Things} ({IoT}),'' \emph{IEEE Access}, vol.~4, pp. 9575--9579,
  2016.

\bibitem{globecom5}
Z.~Mlika and S.~Cherkaoui, ``Massive {IoT} {Access} {With} {NOMA} in {5G}
  {Networks} and {Beyond} {Using} {Online} {Competitiveness} and {Learning},''
  \emph{IEEE Internet of Things Journal}, vol.~8, no.~17, pp. 13\,624--13\,639,
  Sep. 2021.

\bibitem{globecom10}
A.~Taik, B.~Nour, and S.~Cherkaoui, ``Empowering {Prosumer} {Communities} in
  {Smart} {Grid} with {Wireless} {Communications} and {Federated} {Edge}
  {Learning},'' \emph{arXiv:2104.03169 [cs, eess]}, Apr. 2021.

\bibitem{globecom11}
B.~Nour, S.~Cherkaoui, and Z.~Mlika, ``Federated {Learning} and {Proactive}
  {Computation} {Reuse} at the {Edge} of {Smart} {Homes},'' \emph{IEEE
  Transactions on Network Science and Engineering}, pp. 1--1, 2021.

\bibitem{5r}
Z.~A.~E. Houda, A.~Hafid, and L.~Khoukhi, ``{BrainChain} - {A} {Machine}
  learning {Approach} for protecting {Blockchain} applications using {SDN},''
  in \emph{{ICC} 2020 - 2020 {IEEE} {International} {Conference} on
  {Communications} ({ICC})}, Jun. 2020, pp. 1--6.

\bibitem{r4}
M.~H. ur~Rehman, K.~Salah, E.~Damiani, and D.~Svetinovic, ``Towards
  {Blockchain}-{Based} {Reputation}-{Aware} {Federated} {Learning},'' in
  \emph{{IEEE} {INFOCOM} 2020 - {IEEE} {Conference} on {Computer}
  {Communications} {Workshops} ({INFOCOM} {WKSHPS})}, Jul. 2020, pp. 183--188.

\bibitem{r5}
S.~Rathore, Y.~Pan, and J.~H. Park, ``\BIBforeignlanguage{en}{{BlockDeepNet}:
  {A} {Blockchain}-{Based} {Secure} {Deep} {Learning} for {IoT} {Network}},''
  \emph{\BIBforeignlanguage{en}{Sustainability}}, vol.~11, no.~14, p. 3974,
  Jan. 2019.

\bibitem{globecom12}
M.~Lakoju, A.~Javed, O.~Rana, P.~Burnap, S.~T. Atiba, and S.~Cherkaoui,
  ``\BIBforeignlanguage{en}{“{Chatty} {Devices}” and edge-based activity
  classification},'' \emph{\BIBforeignlanguage{en}{Discover Internet of
  Things}}, vol.~1, no.~1, p.~5, Feb. 2021.

\bibitem{r6}
\BIBentryALTinterwordspacing
J.~Weng, J.~Weng, J.~Zhang, M.~Li, Y.~Zhang, and W.~Luo, ``{DeepChain}:
  {Auditable} and {Privacy}-{Preserving} {Deep} {Learning} with
  {Blockchain}-based {Incentive},'' Tech. Rep. 679, 2018. [Online]. Available:
  \url{http://eprint.iacr.org/2018/679}
\BIBentrySTDinterwordspacing

\bibitem{r7}
S.~Otoum, I.~Al~Ridhawi, and H.~T. Mouftah, ``Blockchain-{Supported}
  {Federated} {Learning} for {Trustworthy} {Vehicular} {Networks},'' in
  \emph{{GLOBECOM} 2020 - 2020 {IEEE} {Global} {Communications} {Conference}},
  Dec. 2020, pp. 1--6.

\bibitem{r11}
L.~Feng, Y.~Zhao, S.~Guo, X.~Qiu, W.~Li, and P.~Yu, ``Blockchain-based
  {Asynchronous} {Federated} {Learning} for {Internet} of {Things},''
  \emph{IEEE Transactions on Computers}, pp. 1--1, 2021.

\bibitem{r12}
Y.~Qu, L.~Gao, T.~H. Luan, Y.~Xiang, S.~Yu, B.~Li, and G.~Zheng,
  ``Decentralized {Privacy} {Using} {Blockchain}-{Enabled} {Federated}
  {Learning} in {Fog} {Computing},'' \emph{IEEE Internet of Things Journal},
  vol.~7, no.~6, pp. 5171--5183, Jun. 2020.

\bibitem{r8}
\BIBentryALTinterwordspacing
J.~Kang, Z.~Xiong, D.~Niyato, Y.~Zou, Y.~Zhang, and M.~Guizani, ``Reliable
  {Federated} {Learning} for {Mobile} {Networks},'' \emph{arXiv:1910.06837
  [cs]}, Oct. 2019. [Online]. Available: \url{http://arxiv.org/abs/1910.06837}
\BIBentrySTDinterwordspacing

\bibitem{r9}
Y.~Lu, X.~Huang, K.~Zhang, S.~Maharjan, and Y.~Zhang, ``Blockchain {Empowered}
  {Asynchronous} {Federated} {Learning} for {Secure} {Data} {Sharing} in
  {Internet} of {Vehicles},'' \emph{IEEE Transactions on Vehicular Technology},
  vol.~69, no.~4, pp. 4298--4311, Apr. 2020.

\bibitem{r13}
X.~Huang, R.~Yu, J.~Kang, and Y.~Zhang, ``Distributed {Reputation} {Management}
  for {Secure} and {Efficient} {Vehicular} {Edge} {Computing} and {Networks},''
  \emph{IEEE Access}, vol.~5, pp. 25\,408--25\,420, 2017.

\bibitem{globecom4}
E.~D.~N. Ndih and S.~Cherkaoui, ``\BIBforeignlanguage{en}{Chapter 17 -
  {Simulation} methods, techniques and tools of computer systems and
  networks},'' in \emph{\BIBforeignlanguage{en}{Modeling and {Simulation} of
  {Computer} {Networks} and {Systems}}}, M.~S. Obaidat, P.~Nicopolitidis, and
  F.~Zarai, Eds.\hskip 1em plus 0.5em minus 0.4em\relax Boston: Morgan
  Kaufmann, Jan. 2015, pp. 485--504.

\bibitem{globecom3}
E.~D. Ngangue~Ndih, S.~Cherkaoui, and I.~Dayoub, ``Analytic {Modeling} of the
  {Coexistence} of {IEEE} 802.15.4 and {IEEE} 802.11 in {Saturation}
  {Conditions},'' \emph{IEEE Communications Letters}, vol.~19, no.~11, pp.
  1981--1984, Nov. 2015.

\end{thebibliography}


\end{document}